# POWER ASSISTED TREND FOLLOWING

By Dr Andreas A. Aigner & Walter Schrabmair

*"The trend is your friend"* is a common saying, the difficulty lies in determining if and when you are in a trend. Is the trend strong enough to trade? When does the trend reverse and how are you going to determine this? We will try and answer at least some of these questions here.

In 1978 a book was published entitled *"New Concepts in Technical Trading Systems"* [1] by a real-estate developer turned trader by the name of J. Welles Wilder Jr. After 10 years of losing money he was broke so he needed to make money fast and wrote this book in the space of 6 months [2, 3]. Since no publisher was willing to publish it, he went on to print and publish it himself, selling it for 65$ at the time, which was extraordinarily expensive compared to other technical books. Since he was able to publish one of the chapters in a well-known magazine called "Commodities" (June 1978) [4] he was able to collect orders prior to printing it. Many of the indicators he developed in this book, such as the ATR, ADX, SAR and foremost the RSI, became industry standards and you would be hard-pressed finding a financial application today that doesn't offer all or at least one of these calculations as standard. As a consequence, he has never rewritten or changed any parts of the book and it still sells for exactly 65$ today [5].

In this book J. Welles Wilder is describing a trend following algorithm which he calls *"Volatility System"* (page 21-34), where he derives the calculation of the True Range (TR) and Average True Range (ATR) of a stock. The algorithm simply follows every new high (or low respectively) with a trailing stop-loss based on the most recent calculation of the ATR and flips direction whenever a close price has triggered this trailing stop-loss. It is called a volatility system because the average true range (ATR) expands and contracts with the amount of volatility that is in the stock price. The True Range or volatility in this case is not calculated according to historical volatility or any of its modifications, such as the Parkinson (1980), Garman-Klass (1980), Rogers Satchell (1991) or Yang & Zhang (2000) estimators [6], instead it is defined as the maximum of three differences, the difference between today's high and low, today's high or low and yesterdays close.

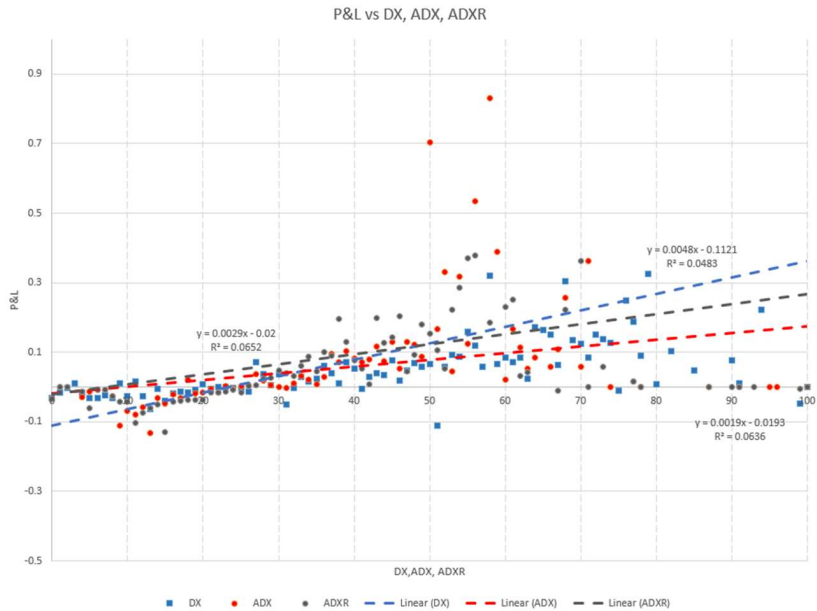

**Figure 1: Plot of P&L vs DX, ADX and ADXR together with a linear regression. $R^2=0.0483$ for DX, $R^2=0.0636$ for ADX and $R^2=0.0652$ for ADXR**

Instead of letting the trend following algorithm stop there, J. Welles Wilder goes several steps further in analyzing the performance of this trend following algorithm. He notes the paradox that *"volatility is always accompanied by movement, but movement is not always accompanied by volatility"* and is deriving methods to attempt to measure and quantify this movement in order to rank each asset and be able to pick the top ranking candidates most suitable for this trend following algorithm. In his book he derives the Volatility Index (VI), the Directional Index (DX), the Average Directional Movement Index (ADX) together with the Average Directional Movement Index Ranking (ADXR) and the Commodity Selection Index (CSI). While these methods do show that the higher the ranking the higher the profit of the trend following algorithm is (Figure 1.) we want to derive a new method to rank assets and indicate when it is a good time to follow the trend or not.

## SECTION 1 – POWER

The Volatility Index trend following algorithm by Wilder has a trailing stop-loss that follows the highest high and lowest low in the trend respectively. The stop-loss is based on the ATR. It follows that if your trend is smaller than the stop-loss you are not going to turn profitable. This is shown in Example 1 on Figure 2 (Ex.1). You start with a Sell and because your stop loss is hit in the next step you reverse position to a Buy only to get stopped out immediately in the next step and in all subsequent steps. This means that this algorithm will not be working when the fluctuations in the stock are equal or smaller than your stop loss limit. For comparison lets look at Example 2 (Ex.2) from Figure 2. Here the trend is 2 times the stop-loss (denoted as 2 ΔF) but still given a fixed period with the same starting and end price as in Example 1 you are going to lose less than in Example 1 (-8ΔF) but still have a loss of -2ΔF. Lets look at another Example 3 (Ex 3) where the trend is 3ΔF long. In this pattern of the trend where we also start and end at the same Price you net 0. Lastly if we look at a trend that is 4ΔF long we get stopped out of our initial Sell and Buy and get stopped out of our Buy 2ΔF higher, where we switch to a Sell. On the way down to reach the exact same finish price we net another 3ΔF such that the sum of all three periods (-ΔF,+2ΔF,+3ΔF) equals 4ΔF.

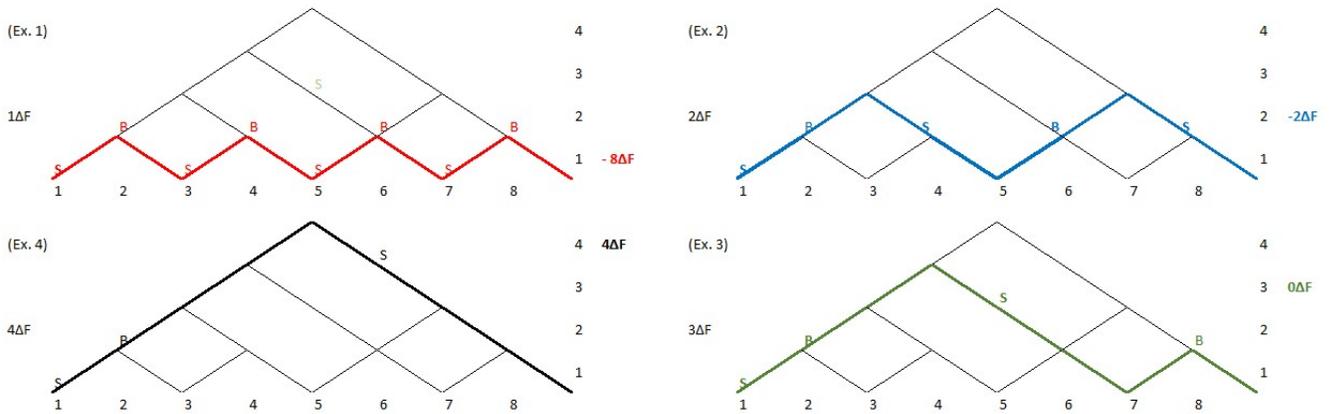

**Figure 2 : Wire graph of example price moves with the same start and end point. ΔF represents the threshold, stop-loss. Showing 4 different amplitudes of the price movement 1-4 times the threshold ΔF.**

We can therefore conclude that the maximum range of a price range has to be a greater multiple of the stop-loss range (N * ΔF) where this multiple should be sufficiently higher than 1 or 2 times.

How can we quantify this maximum range in terms of price? If this trend is simply constant, an absent trend, we know that this quantity should be zero. If the trend is a straight line, a constant trend, then we would want this quantity to be constant.

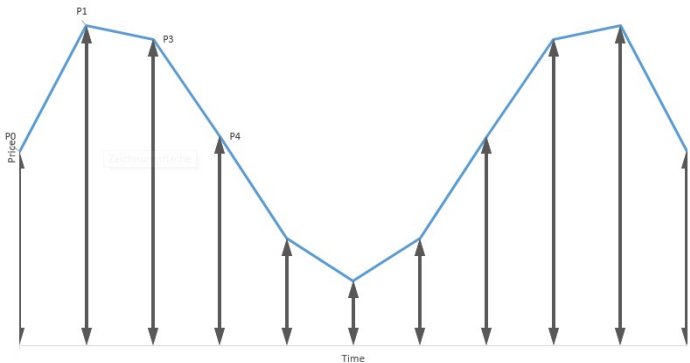

**Figure 3 : A typical signal**

In Physics there is a similar concept called Power, which measures something similar. The Power for periodic signals is defined as the *average energy over a period* [7]. Its formula is

$$Power^N = \frac{1}{N}\sum_{n=0}^{N-1}|P_n|^2 \quad \text{(Eq 1)}$$

Here the absolute value is taken even though it is squared because a signal could be considered complex by nature. For our purposes it can just be dropped. We won't go into discussions about what the Energy of a stock is whether the signal is of infinite or finite length or aperiodic and so forth. We just want to derive an estimate of Power over a range of N values. Furthermore, for this period of N values it makes sense to only look at the changes from the initial value $P_0$ and normalize it. We therefore define a moving window of Power ($Power_j$) at each step in time as such. $P_{n+j}$ being the first value of each moving window.

$$Power_j^N = \frac{1}{N} \sum_{n=0}^{N-1} \left| \frac{P_{n+j}}{P_{0+j}} \right|^2 \qquad \text{(Eq 2)}$$

Having a method to calculate the Power of a stock pattern now, it is only a small step further to define the Power of the Signal and Noise. We calculate the N-Day Moving Average of the stock pattern $MA^N$ and calculate the Power of the Moving Average $MA^N$ which we call the *Signal Power*. Similarly we define the *Power of the Noise* to be the deviation of the Price from the Moving Average (=$P-MA^N$).

$$PowerOfSignal_j^N = \frac{1}{N} \sum_{n=0}^{N-1} \left| \frac{MA_{n+j}^N}{P_{0+j}} \right|^2 \qquad \text{(Eq 3)}$$

$$PowerOfNoise_j^N = \frac{1}{N} \sum_{n=0}^{N-1} \left| \frac{P_{n+j} - MA_{n+j}^N}{P_{0+j}} \right|^2 \qquad \text{(Eq 4)}$$

Since we want to measure the multiple of the stop-loss range we want to calculate the Power of the stop-loss range, the power of the ATR and scale it by the SIC (as defined by Wilder) in order to get a normalized strength of the ATR. In order to get the Power we simply take the square of this value.

$$PowerThreshold_j = \left| \frac{ATR_j}{SIC_j} \right|^2 \qquad \text{(Eq 5)}$$

What we want to plot now are two power ratios. The ratio of the power of noise over the power of the threshold and the ratio of the power of the signal over the power of the threshold. In terms of the signal we are only interested in the excess power so we subtract 1 from the power of the signal. Taking the square root of these two power ratios we arrive at two ratios of the equivalent multiples of the stop-loss range:

$$R_{Signal} = \sqrt{\frac{PowerOfSignal_j^N - 1}{PowerThreshold_j}} \qquad \text{(Eq 6)}$$

$$R_{Noise} = \sqrt{\frac{PowerOfNoise_j^N}{PowerOfNoise_j^N}} \qquad \text{(Eq 7)}$$

# SECTION 2 - EXAMPLES

We are going to take a look at some examples, take for instance a constant value for P, Figure 4. You see the Energy of the Signal and Noise is simply zero after an initial settling down period.

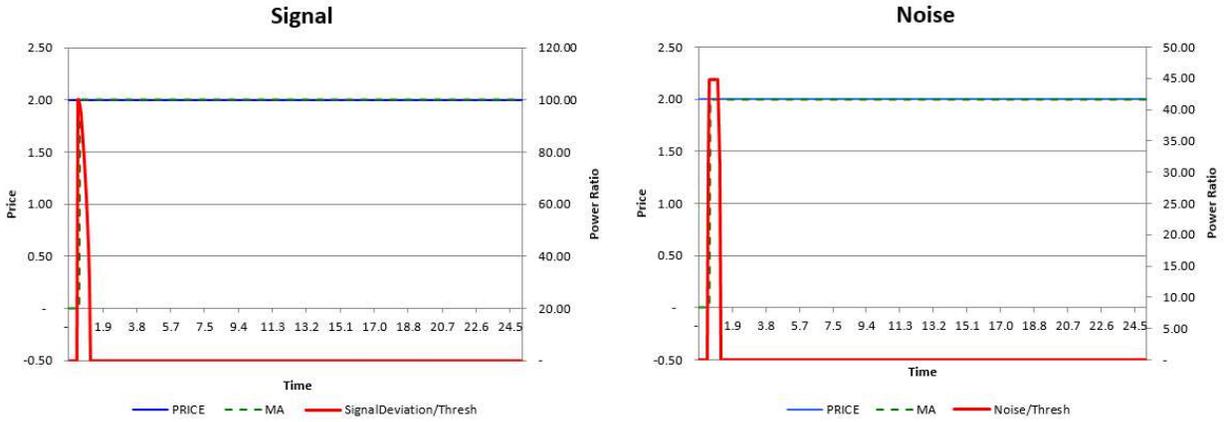

Figure 4 : Plot of the Power Ratio for a constant signal.

Next look at Figure 5 for a linearly increasing Price. The Signal & Noise approaches a constant as you would expect.

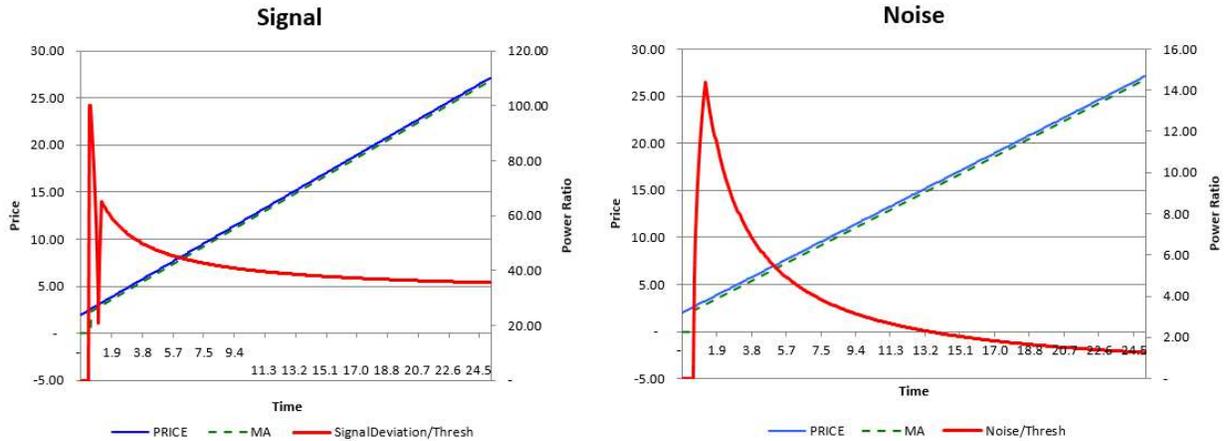

Figure 5 : Plot of the Power Ratio for a linearly increasing signal.

Take a look at a sinusoidal price, you see the Signal and Noise oscillates as well since we are always looking at the instanteous power over a signal of the length N which changes as it moves along.

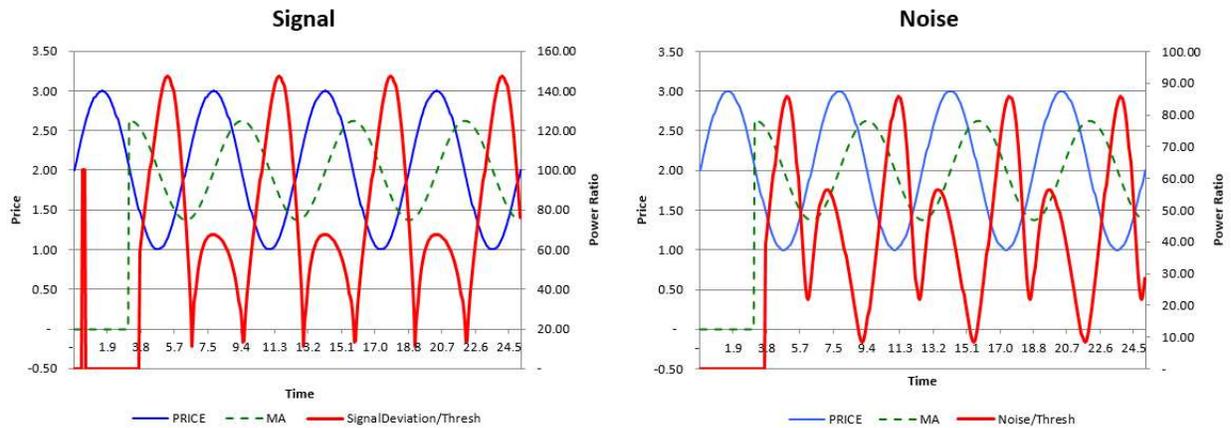

Figure 6 : Plot of the Power Ratio for a sinusoidal signal.

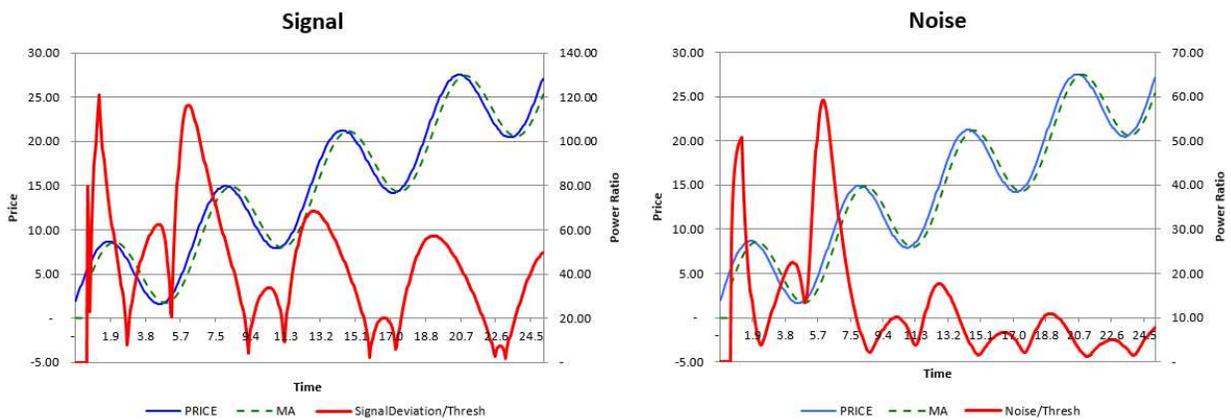

Figure 7 : Plot of the Power Ratio for a combination of sinus and linear functions.

Figure 7 is a chart for a combination of linear and sinusoidal price movement. Figure 8 is a combination of sinus and cosine functions together with a linear function to emulate a stock price movement more accurately. You notice that you get a clear picture of what kind of inherent Power is in the Signal and Noise. Some have quite clear lower bounds others approach these bounds asymptotically. All in all these results show that they are promising in being able to tell us whether we are in a trending mode or not, which is the goal of this indicator.

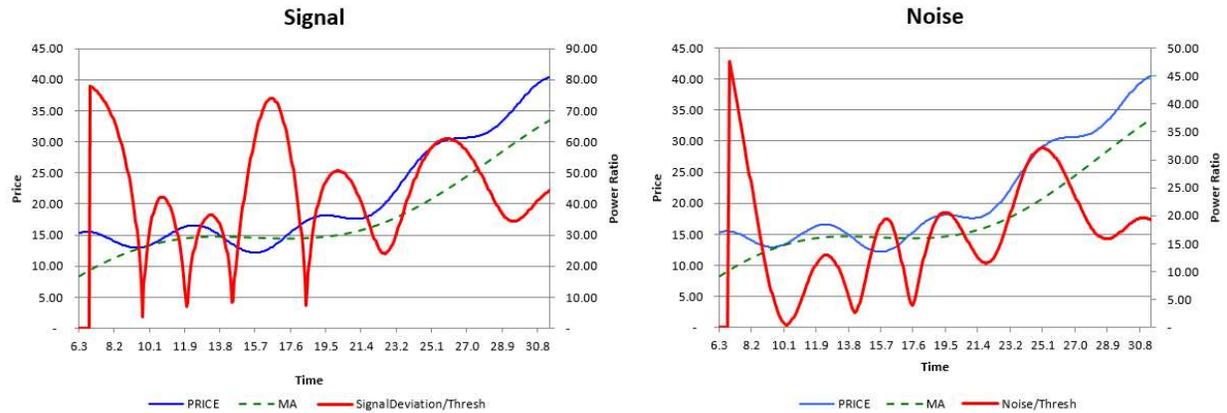

Figure 8 : Plot of the Power Ratio for a combination of sinus, cosine and linear functions.

Next we will chart the Noise and Signal Power Ratios for an example stock TSLA which is strongly trending recently. In Figure 9 we use the 30-day Power and in Figure 10. The 100-day Power. The Noise of the 100-day Power is going to have more energy than the Noise of the 30-day Power calculation since the moving average for the 100-day is a lot smoother than for the 30-day. But both Noise Power calculations show a marked increase in Noise Power post November 2019. While the Signal of the 30-day Power is relatively small prior to November 2019 there is a marked increase post November 2019, since the 30-day moving average is increasing steeply. The Signal Ratio varies between the two ranges since its judging the variation of the moving average. The 30-day moving average takes off steeply after November 2019 but the 100-day moving average still looks relatively flat and the variation of it is relatively mundane. What is more telling in both cases is the increase in Noise Power post November 2019. The Noise of the 100-day increases more than the 30-day since the 30-day Signal already contains most of the power of the pattern. In this sense the 100-day moving average is preferable since it will filter out the longest wavelengths and show the deviation from the moving average as the strongest, which is reflected in the Noise Power Ratio. Figure 11 is a chart of the Power of Threshold which is baked into the Power Ratio calculations. The moves towards the end of the chart are normalized by these extreme levels of thresholds. So even with respect to these the Power of the 100-day is showing a tenfold multiple of the threshold.

Fig 9.

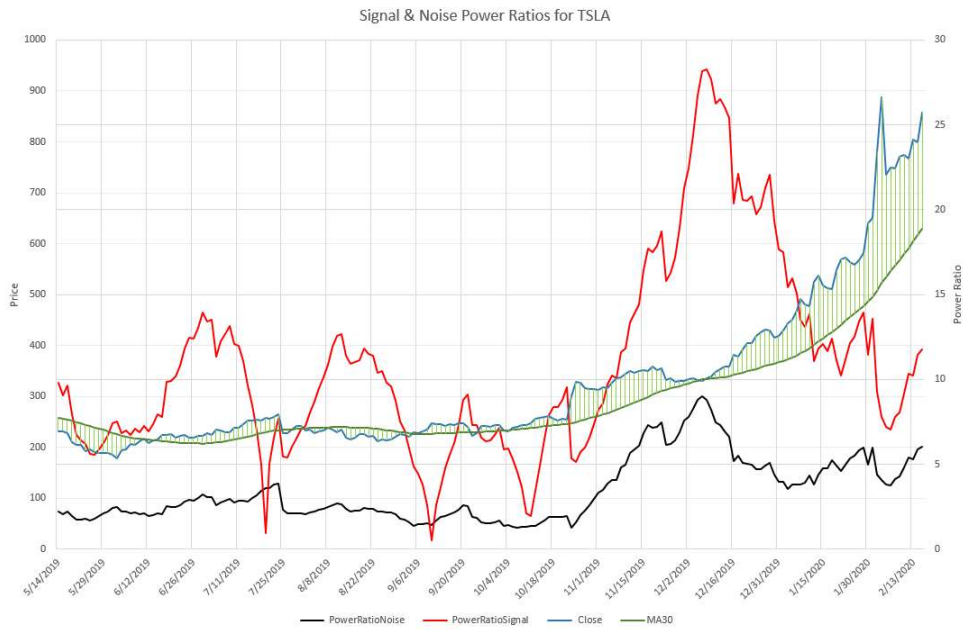

**Figure 9 : Plot of Signal and Noise Power Ratios (30-day) versus the Price of TSLA and its 30-day Moving Average.**

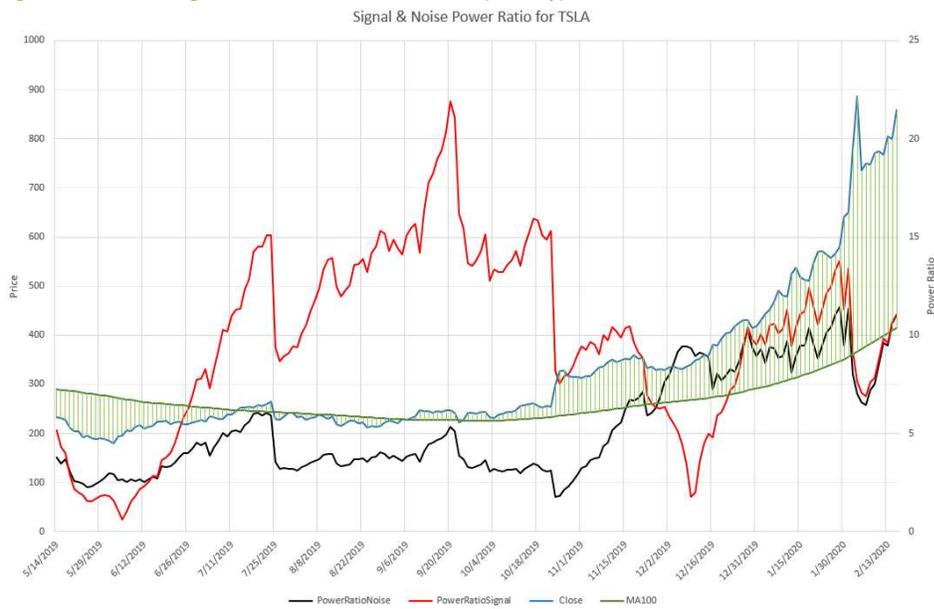

**Figure 10 : Plot of Signal and Noise Power Ratios (100-day) versus the Price of TSLA and its 100-day Moving Average.**

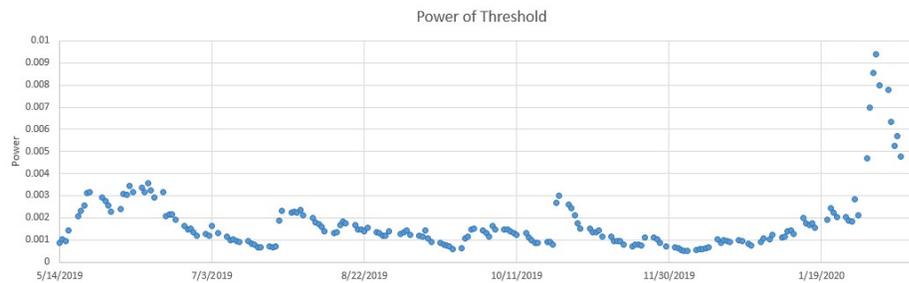

Figure 11 : Plot of the Threshold Power for TSLA

## SECTION 3 – REAL DATA

Now we want to study the performance of Wilder's Volatility Index trend following algorithm combined with the Power Ratios. Wilder suggests using a stop-loss that is 3x the ATR. So he uses a multiple of 3 which he suggests has worked best from his experience. He is actually referring to a range between 2.8 and 3.1. First we want to determine what an ideal value of this multiplier is and we let the multiplier vary from .1 to 8.0. We calculate the PNL across 2700 stocks in the US using a 30-day, 50-day & 100-day range for the Power calculation.

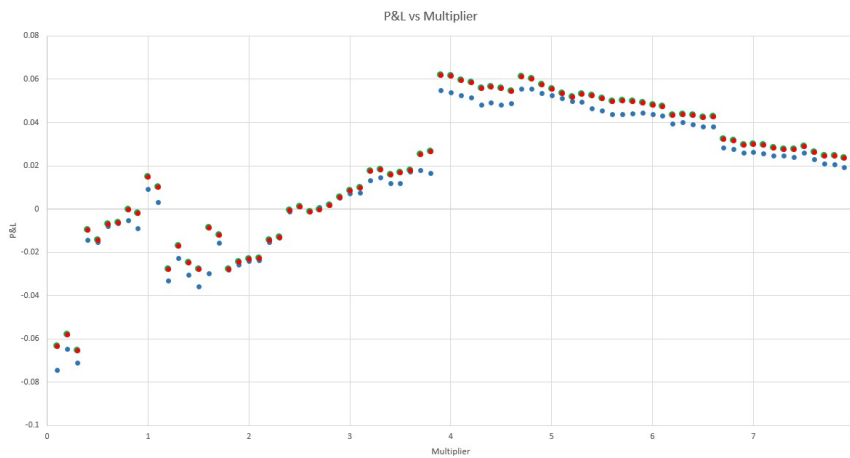

Figure 12 : Plot of P&L versus Multiplier across 2700 US stocks and 30-,50- & 100-day Power calculations.

What we find is that the average P&L across the 2700 names is maximal around a multiplier of 4. We therefore use a multiplier of 4 going forward and we also trigger the trend following only once a level of 4 is breached on the Signal Power Ratio. The following are the results for 3 different ranges (30,50,100) and the plot of P&L versus the Noise Power Ratio and versus the Signal Power Ratio. What is also important to note here that the final Power Ratios that are used here are the final snapshots of these and obviously these vary over time and end up much lower than initially triggered (see Figures 9 & 10). We would expect the relationship between the final Power Ratios and the P&L averages across all 2700 US names to be only a weak one, and a more accurate representation would probably an integral of Power Ratios over time.

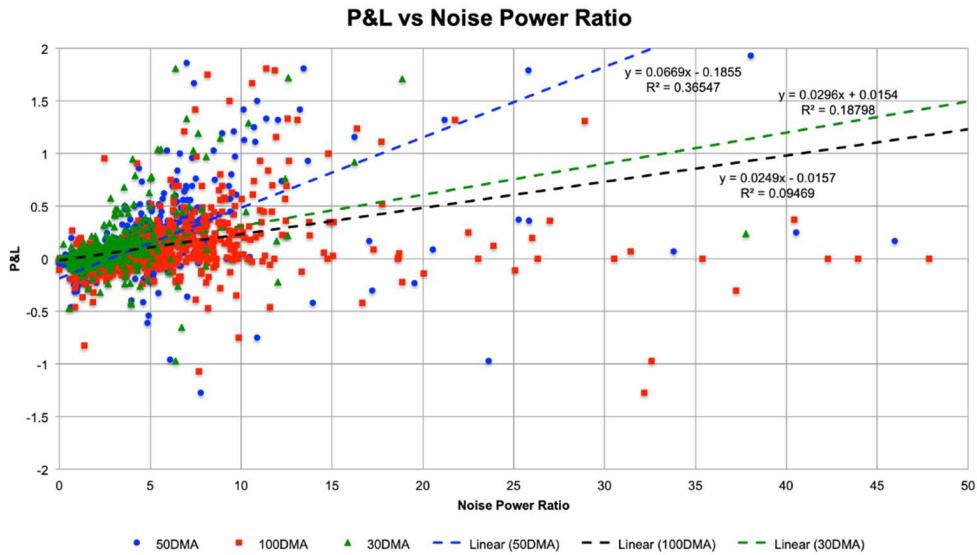

**Figure 13: Plot of P&L versus Noise Power Ratio across 2700 US stocks. . $R^2$=0.37 for 50-day, $R^2$=0.19 for 30-day and $R^2$=0.09 for 100-day Power calculation.**

Figure 13 shows an $R^2$=0.37 for a 50-day Power Ratio and an $R^2$=0.19 for a 30-day Power Ratio. The 100-day Power Ratio has an $R^2$=0.09 for a linear regression of P&L vs the Noise Power Ratio. This is substantially better than the $R^2$=0.05 for the DX, $R^2$=0.07 for the ADXR and the $R^2$=0.06 for the ADX from Wilder (Figure 1).

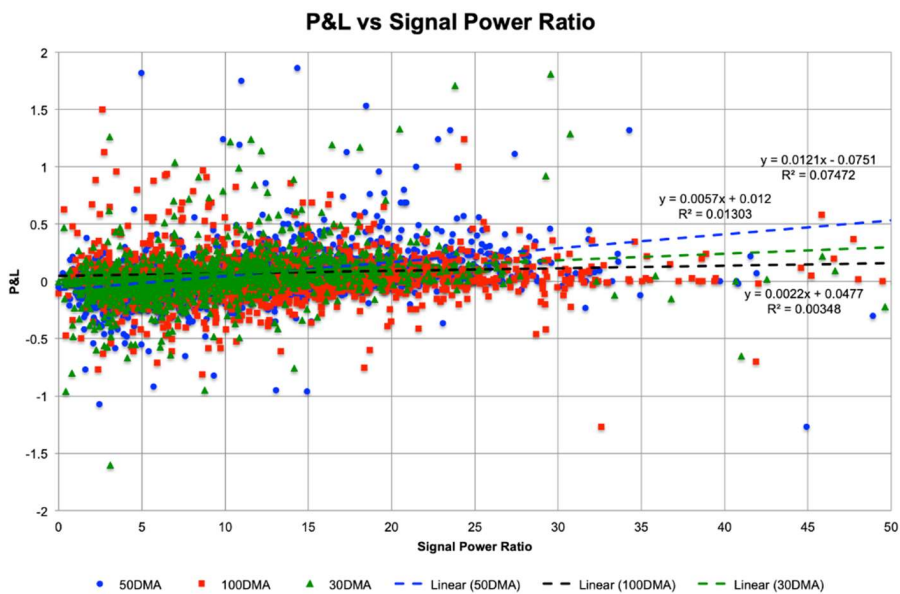

**Figure 14 : Plot of P&L versus Signal Power Ratio across 2700 US stocks. . $R^2$=0.07 for 50-day, $R^2$=0.01 for 30-day and $R^2$=0.003 for 100-day Power calculation**

Figure 14 shows an $R^2$=0.07 for a 50-day Power Ratio and an $R^2$=0.01 for a 30-day Power Ratio. The 100-day Power Ratio has an $R^2$=0.003 for a linear regression of P&L vs the Signal Power Ratio. The Noise Power Ratio has a stronger relationship to P&L than the Signal Power Ratio. We have already deduced this by looking at the example of Figures 9 and 10, but only in a hand waving way. There are may imaginable different ways to make trade entry or exit based on a combination of the Signal and Noise Power Ratio. What we were intending to derive is a measure of trending which assists in entering or exiting a trend following algorithm. Similar to Wilder's Directional Index, Average Directional Index (ADX, ADXR & CSI) we have derived a Power Ratio to assist in ranking trending behavior across assets in a methodical way.

Two interesting parameter sweeps we want to add to this before we conclude are the plot of Power Ratio of Noise and Signal versus a range of Moving Averages and the scatter plot of Noise vs Signal Power Ratio versus the Range.

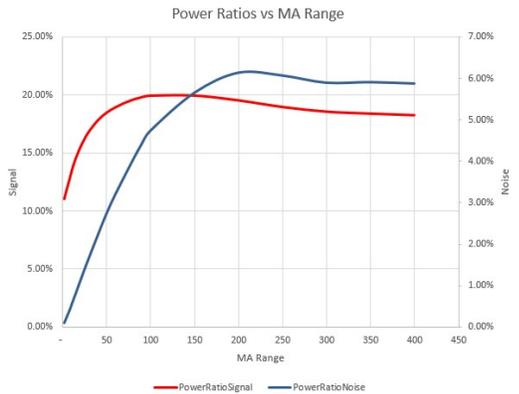

**Figure 15 : Plot of the Signal and Noise Power Ratio versus the MA range from 0 to 400.**

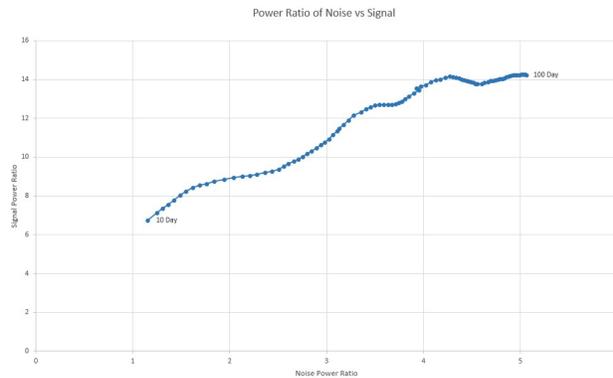

**Figure 16 : Plot of the Signal vs Noise Power Ratio for a range of MA from 10 to 100.**

The charts are suggesting that the Power of Noise and Signal approach a limit as the range increases. Which means that increasing the range to beyond a certain level (ie 100) is not going to significantly change the amount of power measured.

# SECTION 4 – CONCLUSION

The Volatility Index trend following algorithm devised by Wilder in 1978 [1] is essentially a crude version of a machine learning algorithm that indicates if a signal is in a positive or negative trend. It reverses its trend prediction once a threshold is breached which is defined by a multiple of the average day-to-day range (ATR) and the maximum/minimum high/low within the current trend period. Wilder knew that this algorithm needed another indicator to assist with substantiating the trend, since he knew and we have shown above that if the trend is less than a certain multiple of the threshold it will not work as intended and will produce losses. That's why he derived the Directional Movement Index (DX), the Average Directional Movement Index (ADX) and Ranking (ADXR), he also introduced the Commodity Selection Index (CSI) which is essentially a combination of the previous combined with a margin calculation. Here we have derived a calculation of an indicator based on the concepts of energy and power borrowed from discrete signal analysis [7], that we have shown outperforms the indicators developed by Wilder. We have shown this outperformance using a very crude (trigger like) trading methodology assisting the Volatility Index trend following algorithm, which can still be improved upon. More defined entry and exit strategies based off the Power Ratios will likely improve the performance even further. Having said that, the Volatility Index trend following algorithm is an algorithm that many traders (mostly forex traders) use probably without knowing that it can be attributed to the inventor of the RSI, J. Welles Wilder Jr. Many traders use technical indicators to determine entry and exit levels to complement their basic trend following strategy, they 'manually' determine when a stock is in a trend by interpreting the entry and/or exit indicators they use. Here we want to suggest a better method to determine the 'trend' behavior of a stock and call this method therefore the Power Assisted Trend Following.

*Biography: Dr Andreas A. Aigner has a PhD in Mathematics from Monash University, Melbourne, Australia, where he was born and studied. He spent a number of years in Research for various UK universities and worked almost 10 years for Morgan Stanley in Controlling, Trading & Pricing for the Exotic Derivatives desk in Hong Kong. He is now engaged full-time in research and is building a signaling automaton together with his longtime friend and associate Walter Schrabmair, who works at the Medical University of Graz and the Technical University of Graz in various research roles and as a computer whiz. Their contact emails are* andreas@tradeflags.at *and* walter@tradeflags.at